\documentclass[twocolumn,preprintnumbers,amsmath,amssymb]{revtex4}

\usepackage{amsmath}
\usepackage{amsfonts}
\usepackage{amssymb}
\usepackage{graphicx}
\usepackage{dcolumn}
\usepackage{bm}
\usepackage{xspace}
\usepackage{color}
\usepackage{calrsfs}

\DeclareMathAlphabet{\mathpzc}{OT1}{pzc}{m}{it}

\newcommand{\sqrts}{\sqrt{s}}
\newcommand{\sqrtsnn}{\sqrt{s_{_{\rm \textsc{nn}}}}}

\newcommand{\ppbar}{{\rm{p$\bar{\rm p}$}}}
\newcommand{\pA}{{\rm{pA}}}
\newcommand{\pN}{{\rm{pN}}}
\newcommand{\pPb}{{\rm{pPb}}}
\newcommand{\AaAa}{{\rm{AA}}}

\newcommand{\dtwob}{{\rm{d^2b}}}

\newcommand{\sigmaSPS}{\sigma^{{\rm \textsc{sps}}}}

\newcommand{\sigmaeff}{\sigma_{\rm eff,\textsc{dps}}}

\newcommand{\sigmaefftps}{\sigma_{\rm eff,\textsc{tps}}}
\newcommand{\sigmaefftpspA}{\sigma_{\rm eff,\textsc{tps},pA}}

\newcommand{\jpsi}{J/\psi}
\newcommand{\QQbar}    {\rm \textsc{q}\overline{\textsc{q}}}
\providecommand{\ccbar}{\rm c\overline{c}}
\providecommand{\bbbar}{\rm b\overline{b}}

\newcommand*{\eg}{e.g.\@\xspace}
\newcommand*{\ie}{i.e.\@\xspace}
\newcommand*{\cm}{c.m.\@\xspace}

\def\ttt#1{\texttt{\small #1}}
\newcommand{\toppp}{\ttt{Top++}}

\begin{document}

\title{Triple parton scatterings in proton-nucleus collisions at high energies}

\author{David d'Enterria$^1$ and Alexander M.~Snigirev$^2$}
\affiliation{$^1$CERN, EP Department, 1211 Geneva, Switzerland\\
$^2$Skobeltsyn Institute of Nuclear Physics, Lomonosov Moscow State University, 119991, Moscow, Russia}

\date{\today}

\begin{abstract}
A generic expression to compute triple parton scattering (TPS) cross sections in high-energy proton-nucleus (\pA) collisions is derived 
as a function of the corresponding single-parton cross sections and an effective parameter encoding the transverse parton profile of the proton. 
The TPS cross sections are enhanced by a factor of $9\,A\simeq 2000$ in pPb compared to those in proton-nucleon collisions at the same 
center-of-mass energy. Estimates for triple charm ($\ccbar$) and bottom ($\bbbar$) production in pPb collisions at LHC and FCC energies 
are presented based on next-to-next-to-leading order calculations for $\ccbar,\bbbar$ single-parton cross sections. 
At $\sqrtsnn = 8.8$~TeV, about 10\% of the pPb events have three $\ccbar$ pairs produced in separate partonic interactions.
At $\sqrtsnn = 63$~TeV, the pPb cross sections for triple-$\jpsi$ and triple-$\bbbar$ are ${\cal O}$(1--10~mb).
In the most energetic cosmic-ray collisions observed on earth, TPS $\ccbar$-pair cross sections equal the total p-Air inelastic cross section.
\end{abstract}


\maketitle

\section{\label{sec1}Introduction}

The extended nature of hadronic systems and their growing parton density when probed at increasingly higher collision energies,
makes it possible to produce multiple particles with large transverse momentum and/or mass ($\rm \sqrt{p_T^2+m^2}\gtrsim$~3~GeV)
in independent multiparton interactions (MPIs) in high-energy proton-(anti)proton (pp, \ppbar) 
collisions~\cite{Bartalini:2011jp,Abramowicz:2013iva,Bansal:2014paa,Astalos:2015ivw,Proceedings:2016tff}. 
Many experimental final-states --involving the concurrent production of heavy-quarks, quarkonia, jets, and gauge bosons-- 
have been found consistent with double parton scatterings (DPS) processes at Tevatron (see \eg~\cite{Abe:1997xk}) 
and the LHC (see \eg~\cite{Khachatryan:2015pea,Aaboud:2016dea,Aaij:2015wpa} for a selection of the latest results).
Multiple hard parton interaction rates depend chiefly on the transverse overlap of the matter densities of the colliding hadrons, 
and provide valuable information on (i) the badly known 3D parton profile of the proton, (ii) the unknown energy evolution 
of the parton density as a function of impact parameter ($b$), and (iii) the  role of many-parton correlations in the 
hadronic wave functions~\cite{Calucci:2010wg}. In our previous works~\cite{dEnterriaSnigirev}, 
we highlighted the importance of studying DPS also in proton-nucleus (\pA) and nucleus-nucleus (\AaAa) collisions, 
as a complementary means to improve our understanding of hard MPIs in pp collisions. The larger transverse parton density 
in a nucleus (with A nucleons) compared to that of a proton, results in enhanced DPS cross sections coming from interactions 
where the two partons of the nucleus belong to the same or to two different nucleons, providing thereby useful information 
on the underlying multiparton dynamics~\cite{dEnterriaSnigirev,StrikmanEtAl,CattaruzzaTreleani,Blok:2012jr}. 

The possibility of triple parton scatterings (TPS) in hadronic collisions has also been considered in the 
literature~\cite{CalucciTreleani,Maina:2009sj,Snigirev:2016uaq}, and estimates of their expected cross sections 
have been recently provided for pp collisions~\cite{DdESni}. In this paper, we extend our latest work and derive 
for the first time quantitative estimates of the cross sections for observing three separate hard interactions in 
a \pA\ collision 
through a factorized formula which depends on the underlying single-parton scattering (SPS) cross sections normalized by 
the square of an effective cross section $\sigmaefftps$, characterizing the transverse area of triple partonic interactions,
that is closely related to the DPS-equivalent $\sigmaeff$ parameter~\cite{DdESni}.
The paper is organized as follows. In Sec.~\ref{sec:2}, we review the theoretical expression for TPS cross sections 
in generic hadron-hadron collisions --expressed as a convolution of SPS cross sections and generalized parton densities 
dependent on parton fractional momentum $x$, virtuality $Q^2$, and impact parameter $b$-- and its factorized form
as a function of $\sigmaefftps$.
In Section~\ref{sec:3}, a generic expression for 
TPS cross sections in \pA\ collisions is presented based on realistic parametrizations of the nuclear transverse profile. 
As a concrete numerical example, Section~\ref{sec:4} provides estimates for triple charm ($\ccbar$) and bottom ($\bbbar$) cross 
sections from independent parton scatterings in proton-lead (pPb) collisions at the LHC and future circular 
collider (FCC)~\cite{Mangano16b} energies, based on next-to-next-to-leading-order (NNLO) calculations of the 
corresponding SPS cross sections. The main conclusions are summarized in Section~\ref{sec:5}. 

\vspace{-0.5cm}
\section{Triple-parton-scattering cross sections in hadron-hadron collisions}
\label{sec:2}

In a generic hadronic collision, the inclusive TPS cross section from three independent hard parton scatterings ($h h' \to abc$) 
can be written as a convolution of generalized parton distribution functions (PDF) and elementary cross sections summed over 
all involved partons~\cite{CalucciTreleani,Maina:2009sj,Snigirev:2016uaq} 
\begin{eqnarray} 
\label{hardAB}
& &\sigma^{\rm \textsc{tps}}_{hh' \to abc} \nonumber\\  
& & = \frac{\mathpzc{m}}{3!}\,\sum \limits_{i,j,k,l,m,n} \int \Gamma^{ijk}_{h}(x_1, x_2, x_3; {\bf b_1},{\bf b_2}, {\bf b_3}; Q^2_1, Q^2_2, Q^2_3) \nonumber \\
& &\times\hat{\sigma}_a^{il}(x_1, x_1',Q^2_1) \hat{\sigma}_b^{jm}(x_2, x_2',Q^2_2)\hat{\sigma}_c^{kn}(x_3, x_3',Q^2_3)\nonumber\\
& &\times\Gamma^{lmn}_{h'}(x_1', x_2', x_3'; {\bf b_1} - {\bf b},{\bf b_2} - {\bf b},{\bf b_3} - {\bf b}; Q^2_1, Q^2_2, Q^2_3)\nonumber\\
& &\times dx_1 dx_2 dx_3 dx_1' dx_2' dx_3' d^2b_1 d^2b_2 d^2b_3 d^2b.
\end{eqnarray}
Here, $\Gamma^{ijk}_{h}(x_1, x_2, x_3; {\bf b_1},{\bf b_2}, {\bf b_3}; Q^2_1, Q^2_2, Q^2_3)$ are the triple parton distribution functions, 
depending on the momentum fractions $x_1$, $x_2$, $x_3$ at transverse positions ${\bf b_1}$, ${\bf b_2}$, ${\bf b_3}$ of the three partons $i$, $j$, $k$, 
producing final states $a$, $b$, $c$ at energy scales $Q_1$, $Q_2$, $Q_3$, with subprocess cross sections $\hat{\sigma}_a^{il}$, $\hat{\sigma}_b^{jm}$, $\hat{\sigma}_c^{kn}$. 
The combinatorial prefactor $\mathpzc{m}/3!$ takes into the different cases of (indistinguishable or not) final states: $\mathpzc{m}=1$ if $a=b=c$; $\mathpzc{m}=3$ if $a=b$, 
or $a=c$, or $b=c$; and  $\mathpzc{m}=6$ if a, b, c are different. The triple parton distribution functions 
$\Gamma^{ijk}_{h}(x_1, x_2, x_3; {\bf b_1},{\bf b_2}, {\bf b_3}; Q^2_1, Q^2_2, Q^2_3)$ encode all the parton 
structure information of relevance for TPS, and are commonly assumed to be factorizable in terms of longitudinal and transverse components, \ie\
\begin{eqnarray} 
\label{DxF}
& &\Gamma^{ijk}_{h}(x_1, x_2, x_3; {\bf b_1},{\bf b_2}, {\bf b_3}; Q^2_1, Q^2_2, Q^2_3)\nonumber\\
& &= D^{ijk}_h(x_1, x_2, x_3; Q^2_1, Q^2_2, Q^2_3) f({\bf b_1}) f({\bf b_2}) f({\bf b_3}),
\end{eqnarray} 
where $f({\bf b_1})$ describes the transverse parton density of the hadron, often considered a universal function 
for all types of partons, from which the corresponding hadron-hadron overlap function is derived:
\begin{eqnarray} 
\label{f}
T({\bf b}) = \int f({\bf b_1}) f({\bf b_1 -b})d^2b_1 \,.
\end{eqnarray} 
Making the further assumption that the longitudinal components
reduce to the product of independent single PDF, 
$D^{ijk}_h(x_1, x_2, x_3; Q^2_1, Q^2_2, Q^2_3) = D^i_h (x_1; Q^2_1) D^j_h (x_2; Q^2_2) D^k_h (x_3; Q^2_3)$,
the TPS cross section can be expressed in the simple generic form
\begin{equation} 
\label{doubleAB}
\sigma_{hh' \to abc}^{\rm \textsc{tps}} = \left(\frac{\mathpzc{m}}{3!}\right)\, \frac{\sigma_{hh' \to a}^{\rm \textsc{sps}} \cdot
\sigma_{hh' \to b}^{\rm \textsc{sps}} \cdot \sigma_{hh' \to c}^{\rm \textsc{sps}}}{\sigmaefftps^2},
\end{equation} 
\ie\ as a triple product of single inclusive cross sections
\begin{eqnarray} 
\sigmaSPS_{hh' \to a} =  
\sum \limits_{i,k} \int D^{i}_h(x_1; Q^2_1) \,\hat{\sigma}^{ik}_{a}(x_1, x_1') \,D^{k}_{h'}(x_1'; Q^2_1) dx_1 dx_1',
\label{eq:hardS}
\end{eqnarray}
normalized by the square of an effective TPS cross section
\begin{eqnarray} 
\sigmaefftps^2=\left\{\int d^2b \,T^3({\bf b})\right\}^{-1}\,,
\label{eq:sigmaeffTPS}
\end{eqnarray} 
which is closely related to the similar quantity
\begin{eqnarray} 
\label{eq:sigmaeffDPS}
\sigmaeff=\left\{\int d^2b\,T^2({\bf b})\right\}^{-1} \,,
\end{eqnarray}
determined in DPS measurements. 
In the proton-proton case, making use of the expressions (\ref{f}), (\ref{eq:sigmaeffTPS}) and (\ref{eq:sigmaeffDPS}), 
for a wide range of proton transverse parton profiles $f(\bf b)$, we found a simple relationship between the effective 
DPS and TPS cross sections:
\begin{equation}
\label{eq:TPS_DPS_factor}
\sigmaefftps = (0.82\pm 0.11)\cdot\sigmaeff\,,
\end{equation}
which, for the typical $\sigmaeff = 15 \pm 5$ values extracted from a wide range of DPS measurements 
at Tevatron~\cite{Abe:1997xk} and LHC~\cite{Astalos:2015ivw,Abe:1997xk,Khachatryan:2015pea,Aaboud:2016dea,Aaij:2015wpa}, 
translates into
\begin{equation}
\label{eq:TPS_factor}
\sigmaefftps = 12.5 \pm 4.5 \;{\rm mb}\,.
\end{equation}
This data-driven numerical value, together with Eq.~(\ref{doubleAB}), allows the computation of any TPS cross section 
in pp collisions. In the next Section, we extend and exploit these results for the \pA\ case.

\section{Triple-parton-scattering cross sections in proton-nucleus collisions}
\label{sec:3}

The parton flux in \pA\ compared to pp is enhanced by the nucleon number A  and, modulo shadowing effects 
in the nuclear PDF~\cite{eps09}, the single-parton cross section for any hard process is that of proton-nucleon (pN) collisions 
(with $\rm N=p,n$ including their appropriate relative fraction in the nucleus) scaled by the factor A~\cite{d'Enterria:2003qs},
\begin{eqnarray} 
\rm \sigmaSPS_{\rm pA \to a b c} = \sigmaSPS_{\rm pN \to a b c} \int \dtwob \, T_A({\bf b}) = A \cdot \sigmaSPS_{\rm pN \to a b c}\,.
\label{eq:sigmaSPSpA}
\end{eqnarray} 
Here $\rm T_A({\bf b}) = \int f_A(\sqrt{r^2+z^2})dz$ is the nuclear thickness function given by the integral of the 
nuclear parton density function (commonly parametrized in terms of a ``Woods-Saxon'' Fermi-Dirac distribution~\cite{deJager}) 
over the longitudinal direction with respect to the impact parameter ${\bf b}$ between the colliding proton and nucleus, 
normalized to $\rm \int \dtwob \, T_A({\bf b}) = A$. In order to obtain a TPS ``pocket formula'' of the form of Eq.~(\ref{doubleAB})
for \pA\ collisions, 
we follow the approach developed in our previous work for the DPS case~\cite{dEnterriaSnigirev}.
The TPS \pA\ cross section is obtained from the sum of three contributions:
\begin{itemize}
\item A ``pure TPS'' cross section, given by Eq.~(\ref{doubleAB}) for pN collisions scaled by A, namely:
\begin{eqnarray} 
\rm \sigma^{\rm \textsc{tps}, 1}_{\rm pA \to a b c} = A \cdot \sigma^{\rm \textsc{tps}}_{\rm pN \to a b c}\,.
\label{eq:doubleAB1}
\end{eqnarray} 
\item A second contribution, involving interactions of partons from two different nucleons in the
nucleus, depending on the square of $\rm T_{\rm A}$,
\begin{eqnarray} 
\label{eq:doubleAB2}
\sigma^{\rm \textsc{tps}, 2}_{\rm pA \to a b c} = \sigma^{\rm \textsc{tps}}_{\rm pN \to a b c} \cdot 3 \, \frac{\sigmaefftps^2}{\sigmaeff} \, \rm F_{pA},\;\mbox{ with}\\
\rm F_{pA} = \frac{A-1}{A}\int \dtwob\, T_{\rm A}^2({\bf b})\,, 
\label{eq:TpAsq}
\end{eqnarray} 
where the factor (A-1)/A is introduced to account for the difference between the number of nucleon pairs and the number of {\it different} nucleon pairs.
\item A third term, involving interactions among partons from three different nucleons,
depending on the cube of $\rm T_{\rm A}$,
\begin{eqnarray} 
\label{eq:abc3}
\sigma^{\rm \textsc{tps}, 3}_{\rm pA \to a b c} = \sigma^{\rm \textsc{tps}}_{\rm pN \to a b c} \cdot \sigmaefftps^2 \cdot \rm C_{pA}, \;\mbox{ with}\\
\rm C_{pA} = \frac{(A-1)(A-2)}{A^2}\int \dtwob\, T_{\rm A}^3({\bf b})\,.
\label{eq:TpAcub}
\end{eqnarray} 
The factor (A-1)(A-2)/A$^2$ is introduced to take into account the difference between the 
total number of nucleon TPS and that of {\it different} nucleon TPS.
\end{itemize}
The inclusive TPS cross section for three hard parton subprocesses $a$, $b$, and $c$ in \pA\ collisions 
is thus obtained from the sum of the three terms (\ref{eq:doubleAB1}), (\ref{eq:doubleAB2}), and (\ref{eq:abc3}):
\begin{eqnarray} 
\label{eq:triplepA}
\rm \sigma^{\rm \textsc{tps}}_{_{\tiny pA \to a b c}} = A \,\sigma^{\rm \textsc{tps}}_{_{\tiny \rm pN \to a b c}} 
\left[1+3 \, \frac{\sigmaefftps^2}{\sigmaeff} \frac{F_{pA}}{A} + \sigmaefftps^2 \frac{C_{pA}}{A} \right]\,,
\end{eqnarray}
which is enhanced by the factor in parentheses compared to the corresponding TPS cross section in \pN\ collisions scaled by A.
The value of this factor, as well as the relative role of each one of the three TPS components, can be obtained for pPb 
evaluating the integrals (\ref{eq:TpAsq}) and (\ref{eq:TpAcub}) using the standard Fermi-Dirac spatial density 
for the lead nucleus (A~=~208, radius $\rm R_A$~=~6.36~fm, and surface thickness $a$~=~0.54~fm)~\cite{deJager}.
The first integral is identical to the overlap function at zero impact parameter for the corresponding AA collision,
$\rm F_{pA} =(A-1)/A \;{T}_{AA}(0)$~=~30.25~mb$^{-1}$~\cite{dEnterriaSnigirev}. The second one can be obtained 
by means of a Glauber Monte Carlo (MC)~\cite{d'Enterria:2003qs} and amounts to $\rm C_{pA}$~=~4.75~mb$^{-2}$. From the relationship 
(\ref{eq:TPS_DPS_factor}) between effective DPS and TPS cross sections, and the experimental $\sigmaeff = 15 \pm 5$~mb 
value~\cite{Astalos:2015ivw,Abe:1997xk,Khachatryan:2015pea,Aaboud:2016dea,Aaij:2015wpa},
we can finally determine the relative importance for pPb of the three TPS terms of Eq.~(\ref{eq:triplepA}):
$\sigma^{\rm \textsc{tps}, 1}_{\rm pA \to a b c}:\sigma^{\rm \textsc{tps}, 2}_{\rm pA \to a b c}:\sigma^{\rm \textsc{tps}, 3}_{\rm pA \to a b c}
=1:4.54:3.56$. Namely, in pPb collisions, 10\% of the TPS yields come from partonic interactions within just one nucleon 
of the lead nucleus, 50\% involve scatterings within two nucleons, and 40\% come from partonic interactions in three different
Pb nucleons. The sum of the three contributions in Eq.~(\ref{eq:triplepA}) amounts to 9.1, namely 
the TPS cross sections in pPb are nine times larger than the naive expectation based on A-scaling of the corresponding pN TPS cross 
sections, Eq.~(\ref{eq:doubleAB1}). We note that for DPS the equivalent pA enhancement factor was 
$\rm [1+\sigmaeff F_{pA}/A] \simeq 3$~\cite{dEnterriaSnigirev}. The final formula for TPS in proton-nucleus reads 
\begin{equation} 
\label{tripleabc}
\sigma_{\rm pA \to abc}^{\rm \textsc{tps}} =   \left(\frac{\mathpzc{m}}{6}\right)\, \frac{\sigma_{\rm pN \to a}^{\rm \textsc{sps}} \cdot
\sigma_{\rm pN \to b}^{\rm \textsc{sps}} \cdot \sigma_{\rm pN \to c}^{\rm \textsc{sps}}}{\sigma^2_{\rm eff,\textsc{tps}, pA}}\,,
\end{equation} 
where the effective TPS \pA\ cross section in the denominator depends on the effective pp one, and on pure geometric 
quantities directly derivable from the well-known nuclear transverse profile:
\begin{eqnarray} 
\label{eq:sigmaeffpATPS}
\sigma_{\rm eff,\textsc{tps}, pA}^2 & \mkern-10mu = \mkern-10mu & \rm \Big\{A/\sigmaefftps^2 + 2.46\,F_{pA}/\sigmaefftps + C_{pA} \Big\}^{-1}\\
                                   & \mkern-10mu = \mkern-10mu & \rm \Big\{A/156. + F_{pA}[mb^{-1}]/5. + C_{pA}[mb^{-2}]  \Big\}^{-1}\,, \nonumber
\end{eqnarray} 
where the latter equality is obtained using Eqs.~(\ref{eq:TPS_DPS_factor})--(\ref{eq:TPS_factor}).  
The effective TPS cross section in the pPb case amounts thereby to $\sigma_{\rm eff,\textsc{tps}, pA} = 0.29\pm 0.05$~mb.
This value is very robust with respect to the parametrization of the underlying proton and nucleus transverse profiles. Indeed, 
by using simplified Gaussian proton and nucleus transverse densities, all relevant factors in Eq.~(\ref{eq:triplepA}) 
can be analytically calculated, and the effective TPS pA cross section can be simply written as a function of 
the proton and nucleus radii:
$\rm \sigma^2_{\rm eff,\textsc{tps},pA} = 3/4\,\sigmaeff^2/\{A [1+9/2A\,(r_p/R_A)^2 + 4 A^2\,(r_p/R_A)^4]\}$, which amounts 
to $\sigma_{\rm eff,\textsc{tps},pA}\simeq$~0.28~mb (fixing $r_p$ so as to $\sigmaeff = 15$~mb), in perfect agreement with 
our more accurate estimate above.

\section{Triple $\ccbar$ and $\bbbar$ production cross sections in $\rm pA$ collisions}
\label{sec:4} 

\begin{figure*}[htpb!]
\centering
\includegraphics[width=0.99\columnwidth]{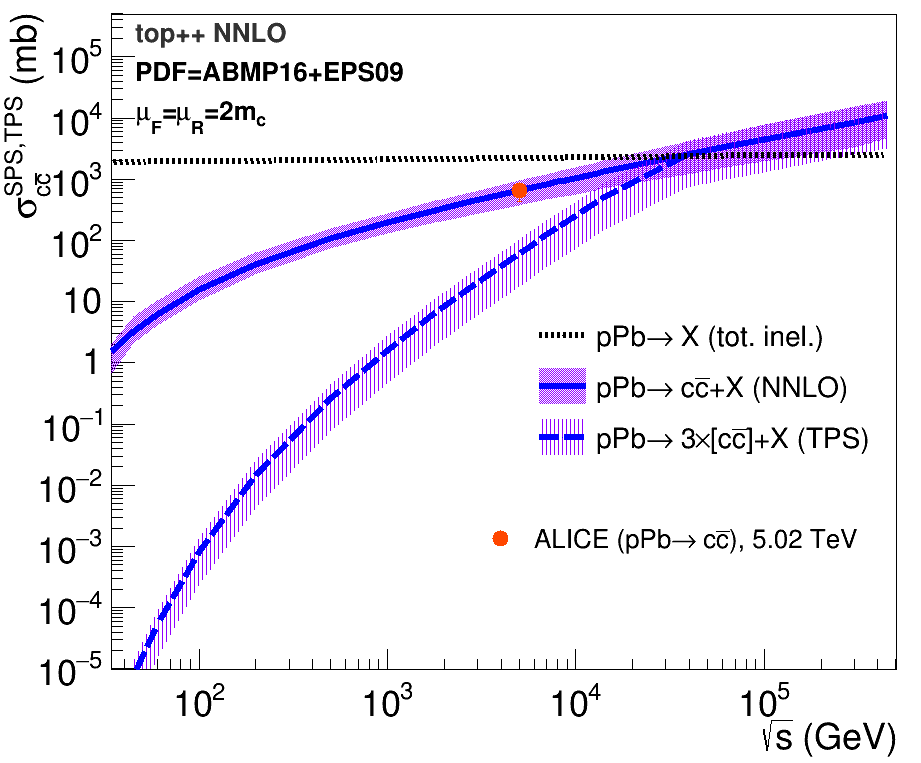}
\includegraphics[width=0.99\columnwidth]{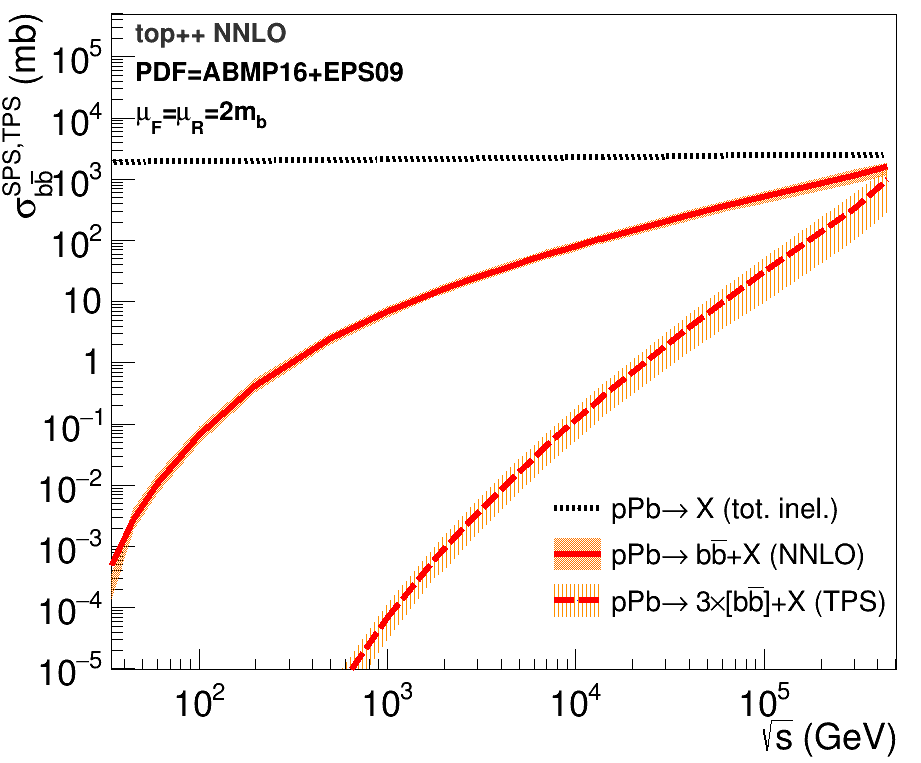}
\caption{Charm (left) and bottom (right) cross sections in pPb collisions as a function of \cm\ energy, in single-parton (solid line) and triple-parton 
(dashed line) parton scatterings, compared to the total inelastic pPb cross section (dotted line). 
Bands around curves indicate scale, PDF (and $\sigmaefftps$, in the TPS case) uncertainties added in quadrature.
The $\rm pPb\to\ccbar+X$ charm data on the left plot has been derived from~\cite{Adam:2016ich}.
\label{fig:1}}
\end{figure*}

As a concrete numerical example of our calculations, following our previous similar pp study~\cite{DdESni}, we compute  
the charm $\rm pPb\to\ccbar+X$ and bottom $\rm pPb\to\bbbar+X$ TPS cross sections first at the LHC and FCC, and then 
also those in proton-air collisions of relevance for ultra-high-energy cosmic rays. These processes are dominated by gluon-gluon 
scattering $gg\to\QQbar$ at low parton fractional momentum $x$, and at high energies the DPS and TPS mechanisms have a 
growing contribution to the total inclusive production. This expectation has been discussed for the DPS case 
in~\cite{Cazaroto:2016nmu}, and we extend those studies to the TPS case here. The TPS heavy-quark cross sections are computed 
via Eq.~(\ref{tripleabc}) for $\mathpzc{m}=1$, \ie\ 
$\sigma_{\rm pPb \to \ccbar,\bbbar}^{\rm \textsc{tps}} = (\sigma_{\rm pN \to \ccbar,\bbbar}^{\rm \textsc{sps}})^3/(6\,\sigma_{\rm eff,\textsc{tps}, pA}^2)$
with $\sigma_{\rm eff,\textsc{tps}, pA}$ given by (\ref{eq:sigmaeffpATPS}), and $\sigma_{\rm pN \to \ccbar,\bbbar}^{\rm \textsc{sps}}$ 
calculated via Eq.~(\ref{eq:hardS}) at NNLO accuracy using a modified version~\cite{DdE} of the $\toppp$ (v2.0) code~\cite{Czakon:2013goa}. 
$\toppp$ is run with $\rm N_f=3,4$ light flavors, charm and bottom pole masses set to $\rm m_{c,b}=1.67, 4.66$~GeV, default renormalization
and factorization scales set to $\rm \mu_R=\mu_F=2\, m_{c,b}$, and using the NNLO ABMP6 PDF of the proton~\cite{Alekhin:2016uxn}
and the nuclear PDF modification factors of the Pb nucleus given by EPS09-NLO~\cite{eps09}. 
The PDF uncertainties include those from the proton and nucleus, as obtained from the corresponding 28 (30) eigenvalues of the ABMP16 
(EPS09) sets, combined in quadrature. The dominant uncertainty is that linked to the theoretical scale choice, which is estimated by 
modifying $\rm \mu_R$ and $\rm \mu_F$ within a factor of two. In the pp case, such a theoretical NNLO setup yields SPS heavy-quark 
cross sections which are larger by up to 20\% at the LHC compared to the NLO~\cite{fonll,mnr} predictions, reaching a better agreement 
with the experimental data~\cite{DdE}, and showing a much reduced scale uncertainty ($\pm50\%,15\%$ for $\ccbar$,$\bbbar$).
In the \pPb\ case, the inclusion of EPS09 nuclear shadowing reduces moderately the total charm and bottom 
cross sections in pN compared to pp collisions, by about 10\% (13\%) and 5\% (10\%) at the LHC (FCC). 
Since the TPS pPb cross section go as the cube of $\sigma_{\rm pN\to\QQbar}^{\rm \textsc{sps}}$, the impact of 
shadowing is amplified and leads to 
15--35\% reductions with respect to the result obtained if one used the pp (instead of the pN) SPS cross section
in Eq.~(\ref{tripleabc}). 
At $\sqrts = 5.02$~TeV, our theoretical SPS prediction ($\sigma_{\rm pPb\to\ccbar}^{\rm \textsc{sps}} = 650 \pm 290_{\rm sc} \pm 60_{\rm pdf}$~mb) 
agrees well with the ALICE total D-meson measurement~\cite{Adam:2016ich} extrapolated using~\cite{fonll} to a total charm cross section 
($\sigma_{\rm pPb\to\ccbar}^{\rm \textsc{alice}} = 640 \pm 60_{\rm stat}\,^{+60}_{-110}\big|_{\rm syst}$~mb, Fig.~\ref{fig:1} left). 

\renewcommand\arraystretch{1.2}%
\begin{table}[htpb]
\caption{\label{tab:1}
Total charm and bottom SPS (NNLO) and TPS cross sections (in mb) in pPb at LHC and FCC
with scales, PDF, and total (quadratically added, including $\sigmaefftps$) uncertainties. The
asterisk indicates that the theoretical prediction of the TPS charm cross section is ``unphysical'' (see text).}
\begin{ruledtabular}
\begin{tabular}{lccc}
 Final state  &  $\sqrtsnn=8.8$ TeV & $\sqrtsnn=63$ TeV \\ 
\colrule
$\sigma(\rm \ccbar+X)$ & $960\pm450_{\rm sc}\pm100_{\rm pdf}$ &  $3400\pm1900_{\rm sc}\pm380_{\rm pdf}$\\
$\sigma(\rm \ccbar\,\ccbar\,\ccbar+X)$ & $200\pm140_{\rm tot}$  &  $8700^*\pm6200_{\rm tot}$ \\
\colrule
$\sigma(\rm \bbbar+X)$ & $72\pm12_{\rm sc}\pm5_{\rm pdf}$ &  $370\pm75_{\rm sc}\pm30_{\rm pdf}$ \\
$\sigma(\rm \bbbar\,\bbbar\,\bbbar+X)$ & $0.084\pm0.045_{\rm tot}$ & $11\pm7_{\rm tot}$ \\
\end{tabular}
\end{ruledtabular}
\end{table}

Table~\ref{tab:1} collects the heavy-quark cross sections and associated uncertainties predicted at the nominal LHC and FCC \cm\ energies.
The large SPS $\ccbar$ cross section 
at the LHC ($\sim$1~b) results in triple-$\ccbar$ cross sections from independent parton scatterings amounting to about 
20\% of the inclusive charm yields. Since the total inelastic pPb cross sections are $\sigma_{\rm pPb}$~=~2.2,~2.4~b at $\sqrtsnn$~=~8.8 and 63~TeV,
charm TPS takes place in about 10\% of the pPb events at 8.8~TeV.
At the FCC, the theoretical TPS charm cross section even overcomes the inclusive charm one. Such an unphysical result indicates 
that quadruple, quintuple,...~parton-parton scatterings are expected to produce extra $\ccbar$ pairs with non-negligible probability
in pPb at $\sqrtsnn = 63$~TeV. The huge TPS $\ccbar$ cross sections at the FCC will make triple-$\jpsi$ production 
observable. Indeed, the SPS $\jpsi$ cross section corresponds to about 5\% of the $\ccbar$ one~\cite{dEnterriaSnigirev}, 
which translates into $\sigma(\rm 3\times\jpsi+X)\approx$~1~mb. Triple-$\bbbar$ cross sections remain comparatively small,
in the 0.1~mb range, at the LHC but reach $\sim$10~mb (\ie\ 3\% of the total inclusive bottom cross section) at the FCC.

Figure~\ref{fig:1} shows pPb cross sections over $\sqrtsnn \approx$~40~GeV--100~TeV for SPS (solid), TPS (dashed) for charm (left) and bottom (right) production, 
and total inelastic (dotted curve, in both plots).  
The TPS cross sections are small at low energies but rise fast with $\sqrts$, as the cube of the SPS cross section evolution. 
Whenever the theoretical central value of the TPS cross section overcomes the inclusive charm cross section, indicative of multiple (beyond three)
$\ccbar$-pair production, we equalize it to the latter. Above $\sqrtsnn \approx$~25~TeV, the total charm and inelastic pPb cross sections are equal
implying that {\it all} \pPb\ interactions produce at least three charm pairs. In the $\bbbar$ case, such a situation only occurs at much higher \cm\ energies, above 500~TeV.

\begin{figure*}[htpb!]
\centering
\includegraphics[width=0.99\columnwidth]{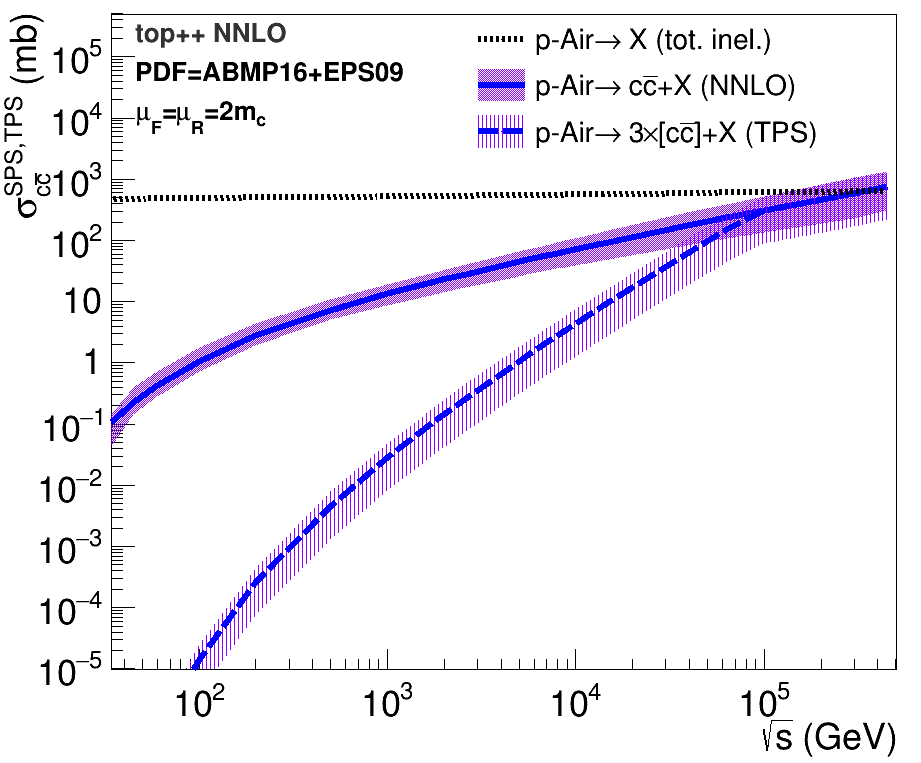}
\includegraphics[width=0.99\columnwidth]{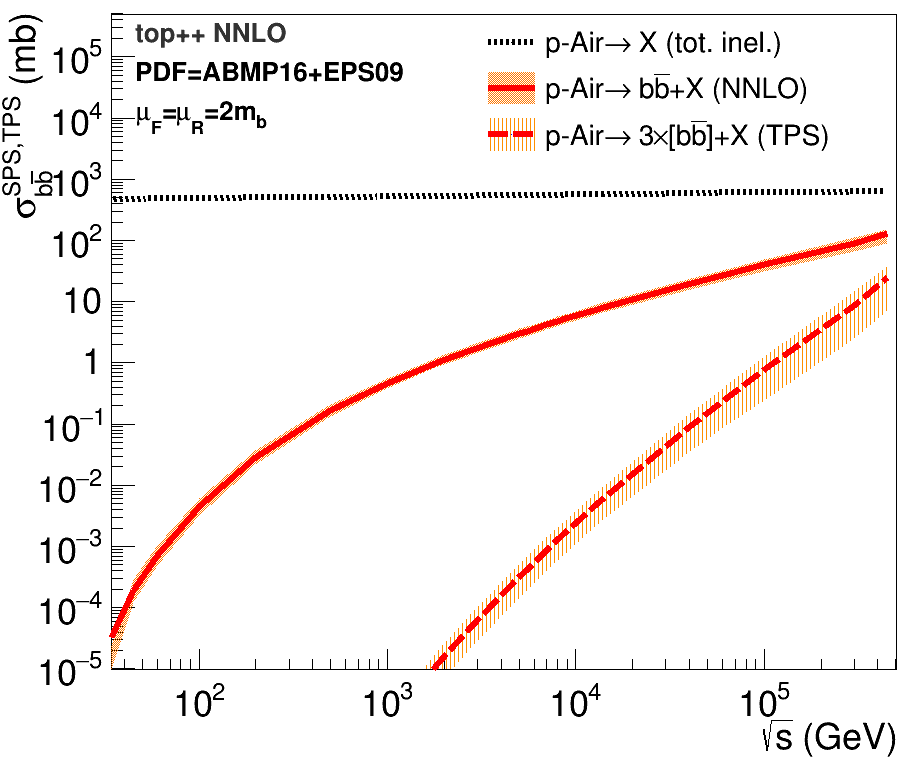}
\caption{Charm (left) and bottom (right) cross sections in p-Air collisions as a function of \cm\ energy, in single-parton (solid line) and triple-parton 
(dashed line) parton scatterings, compared to the total inelastic p-Air cross section (dotted line). 
Bands around curves indicate scale, PDF (and $\sigmaefftps$, in the TPS case) uncertainties added in quadrature.
\label{fig:2}}
\end{figure*}

The most energetic hadronic collisions observed in nature occur in collisions of ${\cal O}\rm (10^{20}\,eV)$ cosmic rays, 
at the so-called ``GZK cutoff''~\cite{Abraham:2008ru}, with N and O nuclei at rest in the upper atmosphere.
To study the amount of triple heavy-quark production produced in such collisions at equivalent \cm\ energies of 
$\sqrtsnn \approx$~430~TeV, we show in Fig.~\ref{fig:2} similar curves as those in Fig.~\ref{fig:1} but for the p-Air case.
The TPS cross sections have been obtained using Eq.~(\ref{tripleabc}) with the same cubic power of the SPS pN cross sections 
computed with the $\toppp$+ABMP6+EPS09 setup, but normalized now to an effective TPS p-Air cross section amounting to 
$\sigma_{\rm eff,\textsc{tps}, pA} = 2.2\pm 0.4$~mb obtained from Eq.~(\ref{eq:sigmaeffpATPS}) using: A~=~14.3 
(from a 78\%-21\% air mixture of $^{14}$N and $^{16}$O), $\rm F_{pA} = 0.51$~mb$^{-1}$, and $\rm C_{pA} = 0.016$~mb$^{-2}$,
the latter two values being obtained via a Glauber MC~\cite{d'Enterria:2003qs}. 
Around the GZK cutoff, the cross section for inclusive as well as TPS charm production equal the total inelastic proton-air 
cross section, $\sigma_{\rm pAir}\approx$~0.61~b, indicating that {\it all} p-Air collisions produce at least
three $\ccbar$-pairs in multiple partonic interactions. In the $\bbbar$ case, about 20\% of the p-Air collisions produce bottom
hadrons but only about 4\% of them have TPS production. These results are clearly of relevance for the hadronic models commonly 
used for the simulation of the interaction of ultrarelativistic cosmic rays with the atmosphere~\cite{dEnterria:2011twh} which, 
so far, do not include any heavy-quark production. Indeed, first, the cosmic ray data~\cite{Aab:2014pza} feature 
unexplained excesses in the number of muons compared to the model predictions, and charmed and bottom hadrons feed more the muonic 
component of the air-shower. In addition, heavy-quark decays are a significant background of high-energy atmospheric 
neutrinos that need to be substracted in searches of astrophysical TeV--PeV $\nu$'s~\cite{Aartsen:2014muf}.
For both reasons, it is worth to explore the impact of such multiple heavy-quark production in the MC generators
commonly used in high-energy cosmic ray and $\nu$ astrophysics.

\section{Summary}
\label{sec:5}

We have derived for the first time estimates of the cross sections for triple parton scattering (TPS)
cross sections in proton-nucleus collisions as a function of the corresponding single-parton cross sections and
an effective $\sigmaefftpspA$ parameter characterizing the transverse density of partons in the proton.
Using NNLO predictions for single heavy-quark production, we have shown that 
three $\ccbar$-pairs 
are produced from separate parton interactions in $\sim$10\% of the pPb events at the LHC. At FCC energies, 
more rare processes such as triple-$\jpsi$ and triple-$\bbbar$ production have cross sections reaching the 1--10~mb range. 
At even higher energies, of a few hundred TeV reachable in the highest-energy collisions of cosmic rays with 
the nuclei in the atmosphere, events producing three charmed hadron pairs occur in all proton-air collisions. 
The quantitative results presented here are of relevance for a proper description and understanding of final states 
with multiple hard particles in heavy-ion collider physics, and for a good control of high-energy $\mu$ and $\nu$ atmospheric 
fluxes in cosmic ray and neutrino astrophysics.\\

{\it Acknowledgments--\;} Discussions with A.P.~Kryukov and M.A. Malyshev on TPS, and with 
M.~Cacciari, M.~Czakon, A.~Mitov and G.~Salam on NNLO heavy-quark calculations are 
gratefully acknowledged.


\end{document}